\documentclass[conference]{IEEEtran}
\IEEEoverridecommandlockouts
\usepackage{bm}
\usepackage{amsmath,amssymb,amsfonts}
\usepackage{algcompatible}
\usepackage[ruled]{algorithm2e}
\usepackage{algpseudocode}
\usepackage{steinmetz}
\usepackage{graphicx}
\graphicspath{{sims/}}	
\usepackage{textcomp}
\usepackage{cite}
\usepackage{color,xcolor}

\usepackage[numbers,sort&compress]{natbib}

\newcommand{\rhant}{
\mbox{$ \; \circ\hspace*{-.5pt}\rule[2pt]{10pt}{.5pt}\!\bullet\; $}}

\newcommand{\Matrix}[1]{{\mathbf{#1}}}

\newcommand{\ive}[1]{{\boldsymbol{#1}}}
\newcommand{\hive}[1]{\hat{\boldsymbol{#1}}}
\newcommand{\hve}[1]{\hat{\mathbf{#1}}}

\newcommand{\ve}[1]{{\mathbf{#1}}}

\newcommand{\Herm}[0]{^{\mathrm{H}}}
\newcommand{\PH}[0]{^{\mathrm{P}}\!}

\newcommand{\diag}[1]{\mbox{diag}\!\left\{#1\right\}}

\newcommand{\ee}{\mathrm{e}}

\newcommand{\bsigma}{\bm{\mathit{\Sigma}}}
\newcommand{\bssigma}{\ve{{\Sigma}}}

\newcommand{\jj}{\mathrm{j}}

\newcommand{\ejk}[1]{\ee^{\jj\Omega_{#1}}}

\DeclareSymbolFont{symbolsC}{U}{ntxsyc}{m}{n}
\SetSymbolFont{symbolsC}{bold}{U}{ntxsyc}{b}{n}
\DeclareMathSymbol{\multimapdotbothB}{\mathrel}{symbolsC}{24}
\newcommand{\dotline}{\multimapdotbothB}
\newif\ifShowCorrections
\ShowCorrectionstrue
\ifShowCorrections
\usepackage[normalem]{ulem}
\definecolor{forgreen}{rgb}{0,0.6,0}
\definecolor{orange}{RGB}{255,140,0}
\definecolor{skyblue}{RGB}{100, 150, 235}

\newcommand{\ikpc}[1]{{\color{forgreen}[#1]}}

\newcommand{\swc}[1]{{\color{forgreen}[#1]}}
\else

\newcommand{\ikpc}[1]{}

\newcommand{\swc}[1]{}
\fi
\algnewcommand\algorithmicfore{\textbf{for}}
\algdef{S}[FOR]{For}[1]{\algorithmicfore\ #1\ \algorithmicdo}

\begin{document}
\renewcommand{\baselinestretch}{0.97}
%%%%%%%%%%%%%%%%%%%%%%%%%%%%%%%%%%%%%%%%%%%%%%%%%%%%%%%%%%%%%%%%%%%%%%%
%
%     TITLE 
%
%%%%%%%%%%%%%%%%%%%%%%%%%%%%%%%%%%%%%%%%%%%%%%%%%%%%%%%%%%%%%%%%%%%%%%%
\title{Least-Squares Khatri-Rao Factorization of a Polynomial Matrix
}
\author{
  \IEEEauthorblockN{Faizan A.~Khattak\IEEEauthorrefmark{1},
    Fazal-E-Asim\IEEEauthorrefmark{2},
    Stephan Weiss\IEEEauthorrefmark{1}, and
    Andr\'{e} L.F.~de Almeida\IEEEauthorrefmark{2}}
  \IEEEauthorblockA{\IEEEauthorrefmark{1}Department of Electronic
    \& Electrical Engineering, University of Strathclyde,
    Glasgow G1 1XW, Scotland}
  \IEEEauthorblockA{\IEEEauthorrefmark{2}Department of
    Teleinformatics Engineering, Federal University
    of Cear\'{a}, Fortaleza, Brazil
    \\\{faizan.khattak,stephan.weiss\}@strath.ac.uk, \{fazalasim,andre\}@gtel.ufc.br}
}

\maketitle

%%%%%%%%%%%%%%%%%%%%%%%%%%%%%%%%%%%%%%%%%%%%%%%%%%%%%%%%%%%%%%%%%
%
%  ABSTRACT
%
%%%%%%%%%%%%%%%%%%%%%%%%%%%%%%%%%%%%%%%%%%%%%%%%%%%%%%%%%%%%%%%%%
\begin{abstract} 
The Khatri-Rao product is extensively used in array processing, tensor decomposition, and multi-way data analysis. Many applications require a least-squares (LS) Khatri-Rao factorization. In broadband sensor array problems, polynomial matrices effectively model frequency-dependent behaviors, necessitating extensions of conventional linear algebra techniques. This paper generalizes LS Khatri-Rao factorization from ordinary to polynomial matrices by applying it to the discrete Fourier transform (DFT) samples of polynomial matrices. Phase coherence across bin-wise Khatri-Rao factors is ensured via a phase-smoothing algorithm. 
The proposed method is validated through broadband angle-of-arrival (AoA) estimation for uniform planar arrays (UPAs), where the steering matrix is a polynomial matrix, which can be represented as a Khatri-Rao product between steering matrix in azimuth and elevation directions. 
\end{abstract}

%%%%%%%%%%%%%%%%%%%%%%%%%%%%%%%%%%%%%%%%%%%%%%%%%%%%%%%%%%%%%%%%%
%
%  I. INTRODUCTION
%
%%%%%%%%%%%%%%%%%%%%%%%%%%%%%%%%%%%%%%%%%%%%%%%%%%%%%%%%%%%%%%%%%
\section{Introduction
   \label{sec:intro}}

The estimation of the angle of arrival (AoA) in narrowband sensor
arrays can be effectively addressed using conventional matrix algebra,
especially the eigenvalue decomposition (EVD)~\cite{book_golub}. Among
the numerous harmonic retrieval algorithms, the spectral MUltiple
SIgnal Classification (MUSIC) algorithm is a widely employed EVD-based
technique~\cite{Schmidt1986,doa_narrowband}.  To circumvent the need
for a grid-based search, its root-MUSIC variant~\cite{rootmusic}
estimates the roots of a polynomial corresponding to the AoA of the
sources impinging on the receiver antenna array.

The root-MUSIC method is particularly suitable for uniform linear
arrays (ULAs), where the Vandermonde structure of the steering vector
can be readily reformulated as a polynomial.  In contrast for uniform
planar arrays (UPAs), where the estimation of both elevation and
azimuth angles is required, applying root-MUSIC directly is
infeasible.  The steering matrix for UPAs can be modelled as the
Kathri-Rao (KR) product of two steering matrices, i.e, horizontal and
vertical steering matrices~\cite{fazal24ITC,krprod}.  Consequently, a
least squares (LS) Khatri-Rao factorization~\cite{krfact,R0} can be
employed to separate the steering matrices, enabling the application
of MUSIC or root-MUSIC independently for elevation and azimuth spatial
angles without any need to pair angles as required in methods
like~\cite{angle_pairing}. The Khatri-Rao factorization has also been utilized in various other signal processing applications, such as in channel and joint channel/symbol estimation for multi-antenna systems 
\cite{R0,R1,R2,R3}, 
array processing~\cite{R4}, 
as well as in individual channel estimation for reconfigurable surfaces~\cite{R5}.

For broadband sensor arrays, the information on the AoA is embedded in
the delay with which signals arrive at the different sensor elements.
Therefore, lag information must be explicitly preserved, and
quantities such as the space-time covariance matrix of the array are
now functions in this lag parameter, and turn into polynomial matrices when
$z$-transformed~\cite{neo23a}. Decompositions such as the polynomial
EVD (PEVD)~\cite{SBR2,SMD,AEEM,AEEM_VT} can lead to generalisations of
narrowband techniques to the broadband case. For ULAs, the MUSIC
algorithm has been generalized to polynomial MUSIC
(PMUSIC)~\cite{pmusic}, capable of performing both spatial and
spectral estimation.  Analogous to the narrowband case, the PMUSIC
approach have been modified to polynomial root-MUSIC
(PRMUSIC)~\cite{prmusic}, significantly lowering the computational
complexity associated with PMUSIC by reducing the number expensive
convolutions that need to be evaluated.

In this paper, we extend the Khatri-Rao factorisation from the narrowband case to that of broadband
multiple-input multiple-output (MIMO) communication systems, and in
particular to polynomial matrices which adequately models broadband MIMO systems.
The proposed approach is DFT domain method where by SVD are computed within the DFT bins and phase smoothing is applied to determine the analytic SVD of a matrix reshaped out from columns of a given polynomial matrix. The DFT size is iteratively increased until convergence metric is satisfied. As an example, the proposed algorithm is applied to an estimated broadband sensor array steering matrix of a UPA for broadband angle of arrival estimation.
%Given the
%fact that PRMUSIC cannot be applied directly to UPAs and the
%computational cost of PMUSIC can be very high due to 2D dense grid
%search, the proposed LS Khatri-Rao factorisation enables the
%decoupling of elevation and azimuth angle estimation process.  This
%enables the application of PRMUSIC directly to 2D AoA estimation, and
%via the PMUSIC at significantly reduced computational cost.  
While
this paper does not focus on quantitatively comparing the
computational cost reductions between PMUSIC and PRMUSIC with a LS
Khatri-Rao factorisation for UPAs, our primary contribution lies in
presenting the formulation and implementation of a LS Khatri-Rao
factorisation for polynomial matrices.

The structure of this paper is as follows: Sec.~\ref{sec:ls_kr}
reviews an LS approximation of an ordinary
matrix by a Khatri-Rao product. In
Sec.~\ref{sec:poly_ls_kr}, this is extended to the case of polynomial
matrices based on an analytic singular value decomposition with its algorithmic implementation outlined in Sec.~\ref{sec:implement}.
A worked example is presented in
Sec.~\ref{sec:example} and for an AoA estimation problem for a
broadband UPA in Sec.~\ref{sec:aoa_upa}. Conclusions are drawn in
Sec.~\ref{sec:concl}.

%%%%%%%%%%%%%%%%%%%%%%%%%%%%%%%%%%%%%%%%%%%%%%%%%%%%%%%%%%%%%%%%%
%
%  II. LS FACTORISATION OF KR PRODUCT
%
%%%%%%%%%%%%%%%%%%%%%%%%%%%%%%%%%%%%%%%%%%%%%%%%%%%%%%%%%%%%%%%%%
\section{Best Least-Squares Approximation of a Matrix by a Khatri-Rao Product
   \label{sec:ls_kr}}

Consider a matrix $\mathbf{A} \in \mathbb{C}^{MN \times P}$ which can be approximated as a Kathri-Rao product of two matrices
$\mathbf{G} = [\mathbf{g}_1, \dotsc, \mathbf{g}_P] \in \mathbb{C}^{M
  \times P}$ and $\mathbf{H} = [\mathbf{h}_1,\dotsc,\mathbf{h}_P] \in
\mathbb{C}^{N \times P}$, defined as
\begin{align}
\label{eq_krfac_ordmat}
\mathbf{A} \approx \mathbf{G} \diamond \mathbf{H} = 
[\mathbf{g}_1 \otimes \mathbf{h}_1 \ \mathbf{g}_2 \otimes \mathbf{h}_2 \ \dots \ \mathbf{g}_P \otimes \mathbf{h}_P],
\end{align}
where $\diamond$ denotes the Khatri-Rao product and $\otimes$ denotes the
Kronecker product.

To determine the factors of the Khatri-Rao product in the LS sense,
one can employ the singular value decomposition (SVD) of matrices
formed from the individual columns of $\mathbf{A} =
[\mathbf{a}_1,\dotsc,\mathbf{a}_P]$. Let $\mathbf{a}_p$ represent the
$p$-th column of $\mathbf{A}$, where $p = 1, 2, \dots, P$. The vector
$\mathbf{a}_p$ can be reshaped into a matrix or tensor using an
unvectorization operation, denoted as
$\ve{A}^{(c)}_p=\text{unvec}\{\mathbf{a}_p\}\in\mathbb{C}^{M\times
  N}$. Using the SVD of $\ve{A}^{(c)}_p=\text{unvec}\{\mathbf{a}_p\}$,
the best rank one approximation can be expressed as
\begin{align}
\label{eq_rank_one_svd}
\ve{A}^{(c)}_p=\text{unvec}\{\mathbf{a}_p\} \approx \mathbf{u}_{p,1}
   \sigma_{p,1} \mathbf{v}_{p,1}\Herm \; ,
\end{align}
where $\mathbf{u}_{p,1}$ and $\mathbf{v}_{p,1}$ are the dominant left
and right singular vectors, respectively, and $\sigma_{p,1}$ is the
corresponding dominant singular value of $\ve{A}_p^{(c)}$.  
The SVD in \eqref{eq_rank_one_svd} can be exploited to estimate $\mathbf{g}_p$ and $\mathbf{h}_p$ but
these factors will be complex-valued scalar ambiguous; to see this, note that
it can be expressed as
\begin{align}
\label{eq_amb_g_h}
\hve{g}_p = \mathbf{u}_1\alpha_p, \quad \hve{h}_p = \mathbf{v}_1^*\beta_p,
\end{align}
where $\alpha_p$ and $\beta_p$ are complex scalar values with
constraint $\alpha_p\beta_p=\sigma_{p,1}$.  The phase factor in this
scaling ambiguity stems from the phase ambiguity of singular
vectors~\cite{book_golub}.  
Among this choice for
$\alpha_p,\beta_p$, we restrict ourselves to
\begin{align}
\label{eq_g_n_h}
\hve{g}_p = \mathbf{u}_1 \sqrt{\sigma_1}, \quad \hve{h}_p = \mathbf{v}_1^* \sqrt{\sigma_1}
\end{align}
for simplicity.  By repeating this process for each column of
$\mathbf{A}$, the matrices $\mathbf{G}$ and $\mathbf{H}$ can be
constructed, resulting in the best LS approximation of $\mathbf{A}$ by
a Khatri-Rao product~\cite{krfact}.

%%%%%%%%%%%%%%%%%%%%%%%%%%%%%%%%%%%%%%%%%%%%%%%%%%%%%%%%%%%%%%%%%
%
%  III. LS-KR FACTORISATION EXTENDED TO POLYNOMIAL MATRICES
%
%%%%%%%%%%%%%%%%%%%%%%%%%%%%%%%%%%%%%%%%%%%%%%%%%%%%%%%%%%%%%%%%%
\section{Extension of LS Khatri-Rao Factorization to Polynomial Matrices
  \label{sec:poly_ls_kr}}

In this section, we describe the extension of the Khatri-Rao factorisation to a
polynomial matrix $\ive{A}(z) = [\ive{a}_1(a),\dotsc,\ive{a}_P(z)]$,
via analytic SVDs of $\mathrm{unvec}\{\ive{a}_p(z)\}$, $p=1,\dotsc,P$.

%%%%%%%%%%%%%%%%%%%%%%%%%%%%%%%%%%%%%%%%%%%%%%%%%%%%%%%%%%%%%%%%%
%  III.A Existence of Analytic SVD
%%%%%%%%%%%%%%%%%%%%%%%%%%%%%%%%%%%%%%%%%%%%%%%%%%%%%%%%%%%%%%%%%
\subsection{Existence of Analytic SVD}

For an analytic polynomial matrix $\ive{A}(z)\in\mathbb{C}^{MN\times
  P}$, we seek an approximation via a
Khatri-Rao product of two analytic polynomial matrices
$\ive{G}(z)\in\mathbb{C}^{M\times P}$ and
$\ive{H}(z)\in\mathbb{C}^{N\times P}$ akin to \eqref{eq_krfac_ordmat}
as
\begin{align}
  \ive{A}(z)\approx\ive{G}(z)\diamond\ive{H}(z)=[\ive{g}_1(z) \otimes
      \ive{h}_1(z) \ \dots \ \ive{g}_P(z) \otimes \ive{h}_P(z)],
\end{align}
each $\ive{A}_p^{(c)}(z)=\mathrm{unvec}\{\ive{a}_p(z)\},m=1,\dots,P$
must admit an analytic SVD, i.e.~it must yield factors that are
analytic and hence can be arbitrarily closely approximated by causal
finite impulse response filters via shifts and
truncations~\cite{AEEM_VT}.  
Such an analytic SVD
exists with few exceptions~\cite{PSVD_exist}; these exceptions are avoided here because the elements of $\ive{A}_p^{(c)}(z)$ are
estimated from finite data~\cite{weiss18b,bakhit_loss}. 
Therefore we can find in all cases a
factorisation
\begin{align}
\label{eq_ana_svd}
\ive{A}_p^{(c)}(z)=\ive{U}_p(z)\bsigma(z)_p\ive{V}_p\PH(z) \; ,
\end{align}    
where $\{\cdot\}\PH$ is a para-Hermitian operator equivalent to
Hermitian conjugate and time-reversal
i.e. $\ive{V}_p\PH(z)=\{\ive{V}_p(1/z^*)\}\Herm$~\cite{vad_2}. The
matrices $\ive{U}_p(z)\in\mathbb{C}^{M\times M}$ and
$\ive{V}_p(z)\in\mathbb{C}^{N\times N}$ hold in their columns the
left- and right-singular vectors, respectively, and
$\bsigma_p(z)\in\mathbb{C}^{M\times N}$ contains analytic singular
values on its diagonal. Both factors in \eqref{eq_ana_svd} often have
infinite order, but due to analyticity, they permit
sufficiently accurate approximation through Laurent
polynomials~\cite{AEEM_VT}. Both left and right singular vectors are
unique up to a common allpass factor, i.e.~given an allpass function
$\phi_m(z)$, $\ive{u}_m(z)\phi_m(z)$ and $\ive{v}_m(z)\phi_m(z)$
remains a valid $m$th left- and right-singular vectors.

%%%%%%%%%%%%%%%%%%%%%%%%%%%%%%%%%%%%%%%%%%%%%%%%%%%%%%%%%%%%%%%%%
%  III.B Analytic LS-KR Factors
%%%%%%%%%%%%%%%%%%%%%%%%%%%%%%%%%%%%%%%%%%%%%%%%%%%%%%%%%%%%%%%%%
\subsection{Analytic Least-Squares Khatri-Rao Factors}

Via the analytic SVD factors of $\ive{A}_p^{(c)}(z)$ in
\eqref{eq_ana_svd}, we can represent the frequency dependent version
of \eqref{eq_g_n_h} as
\begin{align}
  \ive{g}_p(\ejk{})=\ive{u}_{p,1}(\ejk{})\sqrt{\sigma_{p,1}(\ejk{})},
      \label{eqn:gpz} \\
  \ive{h}_p(\ejk{})=\ive{v}^*_{p,1}(\ejk{})\sqrt{\sigma_{p,1}(\ejk{})}.
      \label{eqn:hpz}
\end{align}  
As the analytic singular vectors exist~\cite{PSVD_exist},
the analyticity of the factors in \eqref{eqn:gpz} and \eqref{eqn:hpz}
relies on the whether the square root of the analytic singular values
is analytic.  Due to $\ive{A}(z)$ being estimated from finite data, $\sigma_{p,1}(\ejk{})$ is strictly positive and
$2\pi-$periodic for $p=1,\dots,P$, its square root will also be
$2\pi-$periodic and therefore, analytic~\cite{PSVD_exist,analytic_exist}.  Similar to the analytic SVD
factors, both $\ive{g}_p(z)$ and
$\ive{h}_p(z)$ will generally be of infinite order.
However, due to analyticity, the coefficients will decay at least as
fast as exponential functions, therefore, a finite order Laurent
polynomial approximation can be obtained via delay and truncation
operations akin to analytic eigenvectors case for parahermitian
matrices~\cite{book_ana,AEEM_VT}.

The complex-	valued scalar ambiguity translates into analytic
functions, i.e.~$\alpha_p(z)$ and $\beta(z)_p$ can be constrained such that
$\alpha_p(z)\beta_p(z)=\sigma_{p,1}(z)$. Among the infinite solutions to
this constraint, we here seek
$\alpha_p(z)=\beta_p(z)=\sqrt{\sigma_{p,1}(z)}$ for $p=1,\dots,P$. Note that the
square root of a polynomial can be determined via a Maclaurin
series, with an example given in~\cite{weiss18a}.

%%%%%%%%%%%%%%%%%%%%%%%%%%%%%%%%%%%%%%%%%%%%%%%%%%%%%%%%%%%%%%%%%
%
%  IV. DFT-DOMAIN IMPLEMENTATION
%
%%%%%%%%%%%%%%%%%%%%%%%%%%%%%%%%%%%%%%%%%%%%%%%%%%%%%%%%%%%%%%%%%
\section{DFT-Domain Implementation
  \label{sec:implement}}

%%%%%%%%%%%%%%%%%%%%%%%%%%%%%%%%%%%%%%%%%%%%%%%%%%%%%%%%%%%%%%%%%
%  IV.A SVD in Sample Points
%%%%%%%%%%%%%%%%%%%%%%%%%%%%%%%%%%%%%%%%%%%%%%%%%%%%%%%%%%%%%%%%%
\subsection{SVD of $\ive{A}^{(c)}_p(z)$ in Sample Points}

In order to determine LS factorisation of $\ive{A}(z)$ as a Khatri-Rao
product, 
we may implement the ordinary matrix factorization within the
sample points of $\ive{A}_p^{(c)}(z)$ akin to similar efforts in
generalising polynomial matrix
algebra~\cite{neo23a,bakhit_psvd,AEEM,AEEM_VT}.  The conventional SVD
computed independently in $K$ sample points
$\ive{A}_p^{(c)}(\ejk{k})=\ve{A}_{p,k}^{(c)}$, with
$\Omega_k,k=0,\dots, (K-1)$, produces
\begin{align}
\label{eq_bin_svd}
\ve{A}_{p,k}=\ve{U}_{p,k}\ve{\bssigma}_{p,k}\ve{V}_{p,k} \; ,
\end{align}
where $\ve{U}_{p,k}=[\ve{u}_{p,1,k},\dotsc,\ve{u}_{p,M,k}]$ and
$\ve{V}_{p,k}=[\ve{v}_{p,1,k},\dotsc,\ve{v}_{p,N,k}]$ are unitary
matrices comprising the left- and right-singular vectors,
respectively.
The diagonal matrix
$\ve{\Sigma}_{p,k}=\diag{\sigma_{p,1,k},\dots,\sigma_{p,N,k}}$
contains the bin-wise real-valued singular values such that
$\sigma_{p,i,k}>\sigma_{p,i+1,k}~\forall k, i=1,\dots,(M-1)$.  Both
the left- and right-singular vectors possess a phase ambiguity i.e.,
$\ve{u}_{m,k}\ee^{\jj\theta_{m,k}}$ and
$\ve{v}_{m,k}\ee^{\jj\theta_{m,k}}$ are also a valid $m$th left- and
right-singular vector of $\ve{A}_{p,k}$.  Here we refer to the SVD in
\eqref{eq_bin_svd} as the bin-wise SVD in the $k$th frequency bin.

The bin-wise SVD in the $k$th bin relates to the samples of analytic
SVD at $z=\ejk{k}$~\cite{svd_eusipco24,bakhit_loss} as
\begin{align}
\label{eq_an_bw_sig}
\bsigma_p(\ejk{k})& =\ve{\Sigma}_{p,k} \; ,\\
\ive{U}_p(\ejk{k})& =\ve{U}_{p,k}\ve{\Phi}_{p,k}~, \\
 \label{eq_an_bw_v}
\ive{V}_p(\ejk{k})& =\ve{V}_k\left[ \begin{array}{cc} \ve{\Phi}_{p,k} & \Matrix{0}  \\
    \Matrix{0} &  \ve{\Phi}^\prime_{p,k} \end{array} \right]\; ,
\end{align}
where $\ve{\Phi}_{p,k}$ is diagonal matrix of phase shifts of size $M\times M$ and $\ve{\Phi}'_{p,k}$ is unitary are diagonal unitary matrix of size $(N-M)\times (N-M)$.  As a result, the singular vectors of the bin-wise SVD,
computed within the sample points of $\ive{A}^{(c)}_{p}(z)$ on the
unit circle, suffers from a loss of phase coherence between adjacent
bins~\cite{AEEM_VT}.

%%%%%%%%%%%%%%%%%%%%%%%%%%%%%%%%%%%%%%%%%%%%%%%%%%%%%%%%%%%%%%%%%
%  IV.B Sample Points 
%%%%%%%%%%%%%%%%%%%%%%%%%%%%%%%%%%%%%%%%%%%%%%%%%%%%%%%%%%%%%%%%%
\subsection{Sample Points of $\ive{g}_p(z)$ and $\ive{h}_p(z)$}

The sample points $\ive{g}_p(\ejk{k}),\ive{h}_p(\ejk{k})$ can be
determined via the relation between the sample points of analytic SVD of
$\ive{A}_p^{(c)}(z)$ at $z=\ejk{k}$ and the bin-wise SVD in
\eqref{eq_bin_svd} through \eqref{eq_an_bw_sig}-\eqref{eq_an_bw_v} as
\begin{align}
\ive{g}_p(\ejk{k})&=\ve{u}_{p,1,k}\ee^{\jj\phi_{p,1,k}}\sqrt{\sigma_{p,1,k}}\\
\ive{h}_p(\ejk{k})&=\ve{v}^*_{p,1,k}\ee^{-\jj\phi_{p,1,k}}\sqrt{\sigma_{p,1,k}} \; .
\end{align}

This shows that the loss of phase coherence translates to the sample
points of the columns of the LS Khatri-Rao factors i.e.~$\ive{G}(z)$ and
$\ive{H}(z)$. Therefore, unless the phase coherence is established via
e.g.~phase smoothing~\cite{AEEM_VT,bakhit_psvd}, the solution may not
be analtyic, or at the very least the polynomial orders of the
estimate of $\ive{G}(z)$ and $\ive{H}(z)$ may be significantly larger
than necessary. Note that $\ive{g}_p(z)$ and $\ive{h}_p(z)$ are phase
coupled because both left- and right-singular vectors are
phase-coupled via a common allpass function. Hence, phase smoothing is
only required once for any column index $p$.

\subsection{Algorithm and Convergence Metric}

To estimate $\ive{G}(z)$ and $\ive{H}(z)$, the analytic SVD method
of~\cite{bakhit_psvd} is applied to determine only the dominant left
and right singular vectors i.e., $\ive{u}_{p,1}(z),\ive{v}_{p,1}(z)$
of $\ive{A}_p^{(c)}(z)$ for $p=1,\dots,P$ instead of the GSMD~\cite{GSMD} and the GSBR2~\cite{PSVD_GSBR2} algorithms which provide approximate diagonalization and do not offer reduced SVD computation of a $\ive{A}_p^{(c)}(z)$. 
The analytic SVD method requires
invoking a phase smoothing algorithm~\cite{AEEM_VT} $P$ times in total. 
For the
dominant analytic singular value $\sigma_{p,1}(z)$ of $\ive{A}_p^{(c)}(z),p=1,\dots,P$, it is recommended to use the method in~\cite{AEEM_SC}.  With analytic SVD factors computed for all $P$
polynomial matrices, it remains to determine the square root of the analytic singular values with sufficient accuracy.  To determine the square of para-Hermitian polynomials i.e. estimated analytic singular values, we compute the square in root in the DFT domain
\begin{align}
\label{eq_zeta_1}
\tilde{\sigma}_{p,1}(\ejk{k})=\sqrt{{\sigma}_{p,1}(\ejk{k})},k=0,\dots,(K-1)
\end{align} 
at an increasing
DFT size until the time-domain aliasing becomes negligible. For this,
we define as metric
\begin{align}
\label{eq_zeta_2}
  \zeta_p=\frac{\sum_{\tau}||\tilde{\sigma}^{(K)}_{p,1}[\tau]-
    \tilde{\sigma}^{(K/2)}_{p,1}[\tau]||_\mathrm{F}^2}{\sum_{\tau}
     ||\tilde{\sigma}^{(K)}_{p,1}[\tau]||_{\mathrm{F}}^2},
\end{align}
where $\tilde{\sigma}_{p,1}^{(K)}[\tau]$ and
$\tilde{\sigma}_{p,1}^{(K/2)}[\tau]$ is the time-domain equivalent
obtained through a $K$ and $K/2$ point IDFT applied to
$\{\tilde{\sigma}_{p,1}(\ejk{k}) | k=0,\dotsc,(K-1)\}$ and
$\{\tilde{\sigma}_{p,1}(\ejk{2k}) | k=0,\dotsc,K/2-1\}$, respectively.
A reasonably accurate estimate can be obtained based on the metric
$\zeta_p$ which will decrease as $K$ increases, and will eventually
fall below a suitable threshold $\epsilon>0$. This threshold defines
when the DFT size is sufficiently large to minimise both time-domain
aliasing and truncation errors to an acceptable level.  The entire
procedure for this LS Khatri-Rao factorisation is outlined in
Algorithm~\ref{alg_ZC}.

%------------------------------------------------ 
%-------------- Algorithm
\begin{algorithm}[t]
\caption{LS  Khatri-Rao Factorization of $\ive{A}(z)$}
\label{alg_ZC}
Extract $\{\ive{u}_{p,1}(z),\sigma_{p,1}(z),\ive{v}_{p,1}(z)\}$ for $\ive{A}_p^{(c)},p=1,\dots,M$ via~\cite{bakhit_psvd,AEEM_SC,AEEM}\;
Set $\zeta_p=1$\;
\For{$p=1:P$}
    {
    Set $K$ that exceeds order of $\sigma_{p,1}(z)$\;
    \While{$\zeta_p>\epsilon$}
        {
	  determine $\tilde{\sigma}^{(K/2)}_{p,1}[\tau]$ via \eqref{eq_zeta_1}\;
          determine $\zeta_p$ via \eqref{eq_zeta_2}\;
		
          $\tilde{\sigma}_{p,1}(z)=\sum_\tau \tilde{\sigma}^{(K/2)}_{p,1}[\tau]
                  z^{-\tau}$\;
                $K\gets 2K$\;
        }  
        $\hive{g}_p(z)=\hive{u}_{p,1}(z)\tilde{\sigma}_{p,1}(z)$\;
        $\hive{h}_p(z)=\hive{v}^*_{p,1}(z)\tilde{\sigma}_{p,1}(z)$\;
    }
    
\end{algorithm}

%%%%%%%%%%%%%%%%%%%%%%%%%%%%%%%%%%%%%%%%%%%%%%%%%%%%%%%%%%%%%%%%%
%
%  V. WORKED EXAMPLE
%
%%%%%%%%%%%%%%%%%%%%%%%%%%%%%%%%%%%%%%%%%%%%%%%%%%%%%%%%%%%%%%%%%
\section{Worked Example For LS Khatri-Rao Factorisation
   \label{sec:example}}

In this worked example, we construct a $\ive{A}(z)\in\mathbb{C}^{4\times
  2}$ of polynomial order $2$ from known first order matrices $\ive{G}(z)$ and
$\ive{H}(z)$, given below with elements rounded to two decimals:
\begin{align*}
\ive{G}(z) & =\left[ \begin{matrix}
0.98+0.98\jj & -0.55-0.44\jj\\
-0.27-1.19\jj & -0.10+1.37\jj
\end{matrix}\right]+~~~~~~~~~~~~~\\
& \qquad \qquad  ~~~\left[ \begin{matrix}
-1.38-0.87\jj & -1.89+0.25\jj\\
-0.37+0.09\jj & -2.94+0.41\jj
\end{matrix}\right]z^{-1},\\
%------------------ H(z)
\ive{H}(z) & =\left[ \begin{matrix}
0.65-1.77\jj & 0.81+0.50\jj\\
-0.16+0.44\jj & 0.41+0.68\jj
\end{matrix}\right]+~~~~~~~~~~~~~\\
& \qquad \qquad  ~~~\left[ \begin{matrix}
-1.29-0.40\jj & 0.65-0.37\jj\\
0.96+0.05\jj & -0.41+0.26\jj
\end{matrix}\right]z^{-1}.
\end{align*}
The LS Khatri-Rao factors are estimated via
Algorithm~\ref{alg_ZC}. The singular vectors are extracted via the
phase smoothing algorithm at a DFT size of $K=32$. The
$\tilde{\sigma}_{p,1}(z),p=1,\dots,P$ are determined with a
time-domain aliasing threshold of $\epsilon=10^{-6}$.  Subsequently,
the accuracy of the estimated LS Khatri-Rao factors, i.e.,
$\hive{G}(z)$ and $\hive{H}(z)$, is determined, without applying any
truncation, through a normalised difference w.r.t~$\ive{A}(z)$ as
\begin{align}
\xi=\frac{1}{K}\frac{\sum_{k=1}^{K}\lVert \ive{A}(\ejk{k})-\hive{G}(\ejk{k})\diamond\hive{H}(\ejk{k})\rVert_\mathrm{F}^2}{\sum_{k=1}^{K}\lVert \ive{A}(\ejk{k})\rVert_\mathrm{F}^2},
\end{align}
where $\lVert\cdot\rVert_\mathrm{F}$ denotes the Frobenius norm of a
matrix.

The estimated LS Khatri-Rao factors are of polynomial order is $31$ after truncating trailing zero/coefficients
and the error metric metric $\xi$ is determined at a moderately high
$K=256$ to be of order $10^{-9}$. This worked example shows that the
LS Khatri-Rao factorisation can be applied to polynomial matrices
using the analytic SVD approach via phase smoothing.

%We further determine whether the determined LS-KR solution for $\ive{A}(z)$ have less mismatch for a small perturbation $\ive{E}(z)$ than the Frobenius norm of the perturbation i.e. 
%\begin{align*}
%\lVert\hive{A}(\ejk{})-\ive{X}(\ejk{})\rVert_\mathrm{F}-\lVert\ive{A}(\ejk{})-\ive{X}(\ejk{})\rVert_\mathrm{F}\le\lVert\ive{E}(\ejk{})\rVert_{\mathrm{F}}  
%\end{align*}
%$\forall\Omega$, where $\hive{A}(z)=\ive{A}(z)+\hive{E}(z)$ and $\ive{X}(\ejk{})=\hive{G}(\ejk{k})\diamond\hive{H}(\ejk{k})$.
%We experiment with $\ive{E}$ drawn from complex Gaussian distributionWe experiment with $\ive{E}(z)$ with Frobenius norm equal to $5\%$ of the Frobenius norm of $\ive{A}(z)$, the difference between the     
%%%%%%%%%%%%%%%%%%%%%%%%%%%%%%%%%%%%%%%%%%%%%%%%%%%%%%%%%%%%%%%%%
%
%  VI. BROADBAND AOA FROM UPA STEERING MATRIX
%
%%%%%%%%%%%%%%%%%%%%%%%%%%%%%%%%%%%%%%%%%%%%%%%%%%%%%%%%%%%%%%%%%
\section{Broadband AoA From UPA Steering Matrix
  \label{sec:aoa_upa}}

\begin{figure}[tb]
  \centerline{\includegraphics[width=5.5cm]{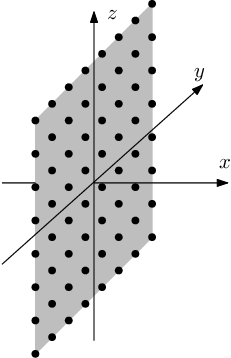}}
  \caption{Uniform planar array with azimuth measured within the
    $xy$-plane against the $x$-axis and the elevation angle measured
    against the $xy$-axis.
    \label{fig:UPA}}
\end{figure}

Returning to the problem of AoA estimation for a broadband planar
array outlined in Sec.~\ref{sec:intro}, we apply the proposed LS
Khatri-Rao factorisation to the broadband steering matrix of a UPA.
We simulate a scenario for two broadband sources firing from
$\{\theta_1,\vartheta_1\}=\{-45^{\mathrm{o}},50^{\mathrm{o}}\}$ and
$\{\theta_2,\vartheta_2\}=\{30^{\mathrm{o}},-45^{\mathrm{o}}\}$ on an
$M\times N$ UPA with $M=N=8$. The azimuth $\theta$ and elevation $\vartheta$ are
the angles of the projection of the vector indicating the propagation
direction into the $xy$ plane, and against the $z$ axis for the
configuration shown in Fig.~\ref{fig:UPA}. Given an ambiguity w.r.t.~a
wavefront either arriving from the front or the back of the array,
i.e.~either from positive or negative $x$-direction, we consider
angular ranges $\theta,\vartheta \in [-\pi, \pi]$.

The broadband steering vectors are constructed based
on windowed sinc function~\cite{windowed_sinc} of order $50$ for $P=2$ sources. The
overall UPA steering matrices are known and can be represented as a
Khatri-Rao product of steering matrices in azimuth and elevation
directions, such that
\begin{align*}
\ive{S}(z)=\underset{\mathbb{C}^{8\times 2}}{\ive{S}_{\mathrm{az}}(z,\theta_{i=1,2})}\diamond\underset{\mathbb{C}^{8\times 2}}{\ive{S}_{\mathrm{el}}(z,\vartheta_{i=1,2})}
\end{align*}
The resultant $\ive{S}(z)\in\mathbb{C}^{64\times2}$ effectively
represents a UPA broadband steering matrix the $P=2$ sources at
$\{\theta_1,\vartheta_1\}=\{-45^{\mathrm{o}},50^{\mathrm{o}}\}$ and
$\{\theta_2,\vartheta_2\}=\{30^{\mathrm{o}},-45^{\mathrm{o}}\}$,
respectively. For the LS Khatri-Rao factorisation,
Algorithm~\ref{alg_ZC} is applied to $\ive{S}(z)$; this generates two
independent steering matrices where the pairing of AoAs are the same
as the LS Khatri-Rao columns. 
Since only two sources are assumed, we
only require the extractio of two rank-one terms, and the phase
smoothing algorithm need to be invoked only twice.

\begin{figure}
   \includegraphics[width=\columnwidth]{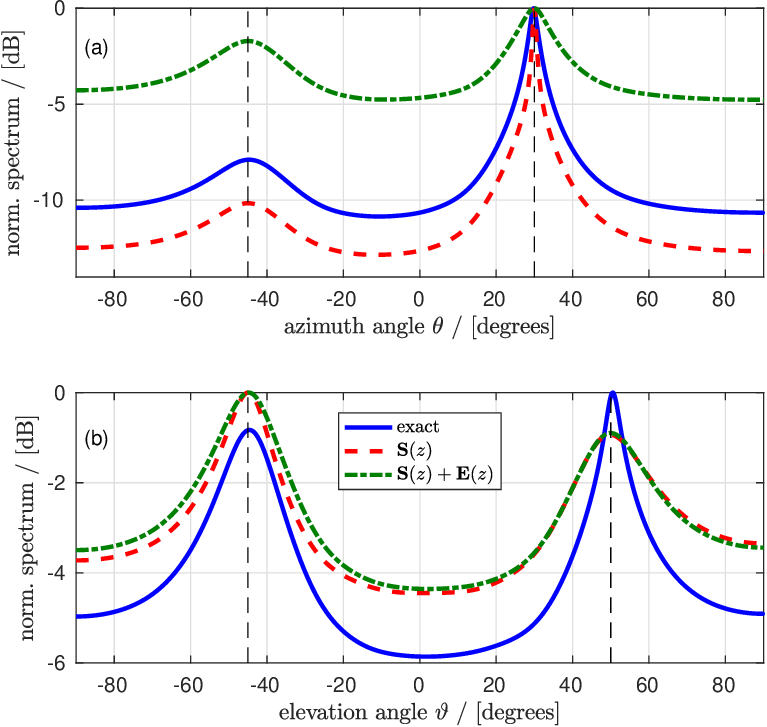}
   \caption{Polynomial spatial MUSIC normalized spectrum for (a)
     azimuth, and (b) elevation directions for ground-truth, LS
     Khatri-Rao factorisation of $\ive{S}(z)$ and LS Khatri-Rao
     factorisation of $\ive{S}(z) + \ive{E}(z) $ perturbed by a term
     $\ive{E}(z)$ with $\sum_n\| \mathbf{E}[n]\|^2_{\mathrm{F}}/(\sum_n\|
     \mathbf{S}[n]\|^2_{\mathrm{F}})=\tfrac{1}{10}$.
    \label{fig_supp}}
\end{figure}

With LS Khatri-Rao factors computed, we can apply the PMUSIC
algorithm because the azimuth and elevation angles are now decoupled.
Hence we capture spatial MUSIC spectrum from the
estimated LS Khatri-Rao factors and compare it against spatial MUSIC
spectrum obtained directly from the ground truth factors
i.e. $\ive{S}_{\mathrm{az}}(z,\theta_{i=1,2})$ ,and
$\ive{S}_{\mathrm{el}}(z,\vartheta_{i=1,2})$.  As shown in
Fig.~\ref{fig_supp}, the spectrum of the estimated LS Khatri-Rao
factors closely aligns with the ground-truth Khatri-Rao factors for
the azimuth and elevation directions, ensuring that the peak
positions, and thus the spatial angles, remain consistent.  Since each
column of $\ive{S}(z)$ is decomposed into Kronecker factors, the angle
pairing ambiguity is inherently resolved, eliminating the need for an
explicit pairing operation as required in prior
methods~\cite{angle_pairing}.

In practice, $\ve{S}[n]\dotline\ive{S}(z)$ is often estimated from
noisy sensor measurements. To model this effect, $\ive{S}(z)$ is
perturbed using $\ive{E}(z)$, whose coefficients are sampled from a
complex Gaussian distribution. The squared Frobenius norm $\mathbf{E}[n]
\rhant \ive{E}(z) = \sum_n \mathbf{E}[n] z^{-n}$ is calibrated to
$\frac{1}{10}$ of squared Frobenius norm of  $\mathbf{S}[n]$.
Subsequently, the LS Khatri-Rao factorisation is applied to the
perturbed $\ive{S}(z)$, and the resulting PMUSIC spectrum is presented
alongside the non-perturbed LS Khatri-Rao factorisation results in
Fig.~\ref{fig_supp}. It is evident from these results that the
estimated AoAs remain very accurate when compared to the
ground-truth AoAs despite the introduced perturbations.

%%%%%%%%%%%%%%%%%%%%%%%%%%%%%%%%%%%%%%%%%%%%%%%%%%%%%%%%%%%%%%%%%
%
%  VII. CONCLUSIONS
%
%%%%%%%%%%%%%%%%%%%%%%%%%%%%%%%%%%%%%%%%%%%%%%%%%%%%%%%%%%%%%%%%%
\section{Conclusion
  \label{sec:concl}}

In this paper, we have introduced the LS Khatri-Rao factorisation of a
polynomial matrix, leveraging the existence of an analytic SVD for
polynomial matrices. By determining the dominant analytic singular
vectors and singular value for the unvectorized representation of each
column of the given polynomial matrix, LS Khatri--Rao factors are
derived in a manner analogous to their counterparts in conventional
matrix analysis.  While we have not yet performed a computational
complexity analysis or comparison for the lack of a suitable
benchmark, we have demonstrated that the proposed algorithm is
particularly applicable to broadband uniform planar arrays (UPAs),
where it enables the decoupling of azimuth and elevation directions
for broadband sources while facilitating automatic angle pairing.

%%%%%%%%%%%%%%%%%%%%%%%%%%%%%%%%%%%%%%%%%%%%%%%%%%%%%%%%%%%%%%%%%
%
%  REFERENCES
%
%%%%%%%%%%%%%%%%%%%%%%%%%%%%%%%%%%%%%%%%%%%%%%%%%%%%%%%%%%%%%%%%%
\bibliographystyle{IEEEtran}
{\footnotesize
%\bibliography{IEEEabrv,bib_total,Stephan_ICC}
}

\end{document}